# Metrics, KPIs, and Taxonomy for Data Valuation and Monetisation - A Systematic Literature Review

EDUARDO VYHMEISTER, University College Cork, Ireland
BASTIEN PIETROPAOLI, University College Cork, Ireland
ALEJANDO MARTINEZ MOLINA, Centro Tecnológico de Investigación, Desarrollo e Innovación en tecnologías de la Información y las Comunicaciones - ITI, Spain
MONTSERRAT GONZÁLEZ-FERREIRO, EGI Fundation, The Netherlands
GABRIEL GONZÁLEZ-CASTANE, Big Data Value Association, Belgium
JORDI ARJONA AROCA, Centro Tecnológico de Investigación, Desarrollo e Innovación en tecnologías de la Información y las Comunicaciones - ITI, Spain
ANDREA VISENTIN, University College Cork, Ireland

Data valuation and data monetisation are complex subjects but essential to most organisations today. Unfortunately, they still lack standard procedures and frameworks for organisations to follow. In this survey, we introduce the reader to the concepts by providing the definitions and the background required to better understand data, monetisation strategies, and finally metrics and KPIs used in these strategies. We have conducted a systematic literature review on metrics and KPIs used in data valuation and monetisation, in every aspect of an organisation's business, and by a variety of stakeholders. We provide an expansive list of such metrics and KPIs with 162 references. We then categorise all the metrics and KPIs found into a large taxonomy, following the Balanced Scorecard (BSC) approach with further subclustering to cover every aspect of an organisation's business. This taxonomy will help every level of data management understand the complex landscape of the domain. We also discuss the difficulty in creating a standard framework for data valuation and data monetisation and the major challenges the domain is currently facing.

CCS Concepts: • **General and reference** → **Surveys and overviews**; **Metrics**.

Additional Key Words and Phrases: Data monetisation, Data valuation, Metrics, Key Performance Indicators, KPIs, Systematic Literature Review

Authors' addresses: Eduardo Vyhmeister, eduardo.vyhmeister@ucc.ie, University College Cork, Cork, Ireland; Bastien Pietropaoli, bpietropaoli@ucc.ie, University College Cork, Cork, Ireland; Alejando Martinez Molina, amartinezmolina@iti.es, Centro Tecnológico de Investigación, Desarrollo e Innovación en tecnologías de la Información y las Comunicaciones - ITI, Valencia, Spain; Montserrat González-Ferreiro, montserrat.gonzalez@egi.eu, EGI Fundation, , The Netherlands; Gabriel González-Castane, gabriel.castane@bdva.eu, Big Data Value Association, Bruxelles, Belgium; Jordi Arjona Aroca, jarjona@iti.es, Centro Tecnológico de Investigación, Desarrollo e Innovación en tecnologías de la Información y las Comunicaciones - ITI, Valencia, Spain; Andrea Visentin, andrea.visentin@ucc.ie, University College Cork, Cork, Ireland.

# 1 INTRODUCTION

With the digitisation of information and the digital transformation of organisations, data has become a critical asset across all sectors [113, 127, 139]. Companies increasingly invest in data collection, storage, and analytics, recognising high-quality data as a key driver of competitive advantage.

The global value of data is estimated in the hundreds of billions of euros. In 2023 alone, estimates reached approximately €350B for the United States, €82B for the EU, €53B for Japan, and €50B for China, with the global market growing rapidly [39]. Despite this surge in value, financially evaluating the gain obtained through the inclusion of data in products and processes remains a complex challenge for many organisations [106, 139]. Data montetisation, as a relatively recent academic topic, lacks a standard definition and is subject to diverse interpretations [83, 97, 110, 113, 139] but common distinctions include: 1) *Internal data monetisation*, which involves modifying internal decision-making and operational efficiency to achieve benefits including monetary gains (e.g., cost optimisation in manufacturing); 2) *Indirect data monetisation*, where data enhances products and services (e.g., recommendation systems); and 3) *Direct external monetisation*, involving the sale of data (e.g., by data brokers).

Data valuation is a concept intrinsically linked to data monetisation. The value of data is often relative and highly dependent on the perspective; combined with the wide array of approaches to monetising data, organisations often find that assigning a "fair" value to data can be a cumbersome task [106]. The absence of a formal valuation framework leads to fragmented practices and inconsistent outcomes. As such, valuing and monetising datasets remain pressing issues in need of robust, standardised solutions.

Whether it is for data monetisation or data valuation, metrics and key performance indicators (KPIs) on the data itself are usually measured. For example, data quality metrics such as accuracy, completeness, or timeliness have been used extensively to evaluate the worth of data and many data quality frameworks exists [104, 105, 132]. However, data quality represents only one aspect of an organisation's data strategy. Other aspects such as the costs (e.g. of acquisition, or storage), the accessibility, the retrievability in case of failure, the fairness and lack of bias in the gathered data, are all examples of metrics which may play a role in the valuation and monetisation of data depending on the chosen strategy and the stakeholders.

The goal of the present study is thus defined as follows:

> **To provide an understanding of different approaches, indicators, and metrics to measure the value (monetary or not) of data. The previous indicators should primarily be focused on the strategies, environments, and stakeholders that participate in the market now and in the foreseeable future.**

To achieve this goal, the following research questions (RQ) need to be answered:

- RQ1 - What are the main Key Performance Indicators (KPIs) and metrics covered in the literature related to data monetisation/valuation?
- RQ2 - What are the main trends and relations observed between the observed KPIs and metrics for data monetisation/valuation?

The main contribution of this paper is twofold: 1) we offer an expansive systematic literature review on metrics and KPIs used for data valuation and monetisation, 2) we propose a new taxonomy for these metrics and KPIs based on the Balanced Scorecard (BSC) [76], hence covering every aspect of a business and catering to various stakeholders.

This paper is organised as follows: Section 2 provides key definitions and background to help understand this study; Section 3 outlines the literature review methodology and the results of our survey; Section 4 describes the taxonomy development process and its high-level structure; in

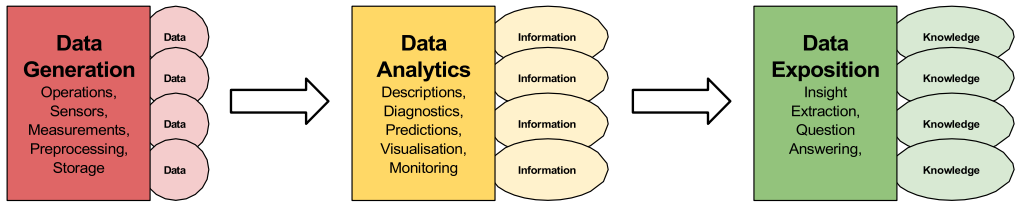

Fig. 1. Data is collected through operations, sensors, and measurements. It is pre-processed and stored, then analysed to extract information through descriptive, diagnostic, or predictive methods. This information, when synthesised and contextualised, yields knowledge in the form of insights and answers [110, 127].

Section 5, we offer our insights about the domain and discuss future challenges and work; finally, Section 6 concludes the paper.

## 2 DEFINITIONS AND BACKGROUND

In this section, we present key definitions and background required to understand this study. In particular, because of a lack of standard definitions [110, 139], we explicit which definitions we use for this study.

### 2.1 Data vs Information vs Knowledge

The terms data, information, and knowledge are frequently discussed in the literature, each reflecting different levels of data processing [1, 48, 107, 145]. *Data* typically refers to raw or minimally processed values (e.g., sensor outputs), *information* is processed data (e.g., summaries or statistics), and *knowledge* represents actionable insights derived from analysed information (e.g., answers to questions). Figure 1 illustrates the difference between the three concepts.

While the term *data* nowadays encapsulates the three concepts, the difference in definition matters in business valuation and monetisation strategies. Therefore, in this study, we use *data* (singular, the concept) as the overall term encapsulating everything, data (plural of datum) for raw and pre-processed data, information for processed data, and knowledge for insights/answers.

### 2.2 Data Value vs Data Price

Data value has a wide range of definition depending on the context or domain in which it is used [13] but, in simple terms, *data value* refers to the perceived worth of data (monetary or otherwise), while *data price* denotes the monetary cost assigned to it (e.g., the selling price of a dataset). Crucially, data value is inherently subjective: the same dataset may be valued differently by different stakeholders or organisations depending on context, use case, and strategic priorities. Data value can be understood as an internal or private estimate of a dataset's utility or impact within an organisation. In contrast, data price is the publicly declared cost of data in a transactional context, such as a data marketplace. As with any asset, an organisation will consider purchasing data when its internal valuation exceeds the seller's price.

### 2.3 Data Valuation vs Data Monetisation

*Data valuation* refers to estimating the worth of data, whether monetary or not. In contrast, *data monetisation*, according to Gartner, "refers to the process of using data to obtain quantifiable economic benefit" [56]. While valuation is typically an internal exercise, monetisation may involve internal use, indirect enhancement of offerings, or direct commercialisation.

## 2.4 Strategies

A business strategy defines an organisation's long-term plan to achieve sustainable competitive advantage. It guides how resources are allocated, priorities are set, and decisions are made to achieve strategic goals aligned with the company's mission and market position. For example, a company focused on sustainability might pursue eco-friendly innovations and operational practices. This focus differentiates the organisation, as well as defines its strategic goals and the kinds of performance indicators it will prioritise to measure progress. With a well-defined strategy, an organisation can establish specific objectives that support its overarching vision.

Strategic direction also shapes how organisations leverage their data. Within the context of data monetisation, the literature identifies three broad categories: internal monetisation, indirect external monetisation, and direct external monetisation [83, 97, 110, 113, 139]. These categories reflect how data is used to generate value - whether through improved internal processes, enhanced services, or direct data sales.

Authors in [110] provide a framework in which they identify various types of data monetisation, the services, the revenue models, and the actors/stakeholders in that market. Report [83] identifies data monetisation strategies based on the data flow, external and internal business strategy and cash flow. The 12 strategies they describe are summarised in table 1. These 12 strategies appear to comprehensively cover those described across the literature [48, 110, 113, 127, 139, 145].

Table 1. Summary of data monetisation strategies, adapted from [83].

| Strategy | Monetisation | Core Idea | Key Requirements |
| --- | --- | --- | --- |
| Reduce Costs | Internal | Use data internally to cut costs — no external sales. | Data culture, cost mapping, KPI tracking. |
| Data Network | Internal | Connect stakeholders (e.g., partners, peers) to exchange data. | Trust, secure sharing protocols. |
| Wrapping | Indirect External | Add data-driven features to products to increase value. | Clear link between product and data layer. |
| Servitisation | Indirect External | Shift from one-off sales to ongoing services (e.g., predictive maintenance). | Smart hardware, analytics, recurring revenue model. |
| New Products or Services | Indirect External | Launch new offerings based on internal data. | Innovation, product/dev capacity, market insight. |
| Data Bartering | Indirect External | Trade data for tools, marketing, or services. | Ownership clarity, legal safeguards, win-win setup. |
| Platform Providing | Indirect External | Host and sell data from multiple sources. | Secure integration, user-friendly platform. |
| Platform Refining | Direct External | Improve external data (e.g., clean, enrich) before resale. | Curation, metadata, quality standards. |
| Data-as-a-Service (DaaS) | Direct External | Sell raw datasets or streams to third parties. | Volume, anonymisation, delivery infrastructure. |
| Information-as-a-Service (IaaS) | Direct External | Provide structured insights or reports. | Analytics skills, visualisation, domain expertise. |
| Answer-as-a-Service (AaaS) | Direct External | Deliver actionable recommendations or dashboards. | Advanced analytics, decision tools. |
| Share for Free | Non-monetary | Publish data to attract users or build ecosystems. | Anonymisation, clear terms of use. |

With a well-defined strategy, an organisation can establish specific objectives that support its overarching vision. These objectives are often broken down into concrete, measurable targets, enabling the company to monitor its trajectory over time. At this stage, KPIs and metrics become essential: they are tools to quantify the strategy and translate it into actionable insights.

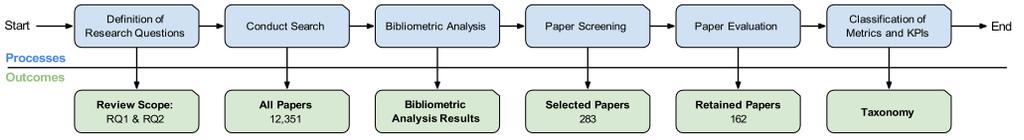

Fig. 2. Systematic literature review processes (top) and outcomes (boNom).

## 2.5 Metrics vs KPIs

While metrics and Key Performance Indicators (KPIs) both measure organisational performance, they differ in scope and strategic relevance.

*Metrics* are broad measurements used in day-to-day operations. While not always strategic, they provide the foundation for performance monitoring and tactical decision-making. Metrics provide detailed insights that help companies evaluate their performance, identify trends, and make improvements in specific areas. They can be objective (e.g., a sensor measure) or subjective (e.g., a human-reported estimate), and qualitative (e.g., the clarity of a dataset) or quantitative (e.g., the number of downloads of a dataset) [13].

On the other hand, *KPIs* are high-level indicators tied to strategic goals. They help assess whether an organisation is on track to achieve its objectives. KPIs are typically derived from operational metrics, constrained by targets, benchmarks, or thresholds. For instance, a metric $M$ can become a KPI when formulated as:

$$KPI_M = \frac{M_{\text{current}} - M_{\text{start}}}{M_{\text{target}} - M_{\text{start}}} \qquad (1)$$

where $M_{current}$ represents the current value of the metric, $M_{start}$ represents the value of the metric when the KPI was set, and $M_{target}$ represents the metric value set as a target by the organisation. If $KPI_M < 0$, then the organisation's performance has degraded; if $0 \leq KPI_M \leq 1$, then the KPI indicates the completion percentage of the set target; if $KPI_M > 1$, then the organisation has over-performed compared to its set target.

KPIs are selected with strategic alignment in mind, focusing only on the aspects of the business that drive growth and impact the organisation's overall objectives, hence they usually are in fewer numbers than metrics. KPIs must be regularly reviewed and updated to remain relevant. In particular, their associated targets should be clearly time-framed (e.g., quarterly, annually) to reflect evolving business goals and ensure accountability. For instance, if a company's growth strategy emphasises customer satisfaction, a relevant KPI might be "achieve a X% customer satisfaction rate by the end of the year." This KPI provides a measurable target and a clear focus area and time frame for the organisation, steering decision-making across teams toward a shared outcome.

## 3 SYSTEMATIC LITERATURE REVIEW

Figure 2 illustrates our research methodology; processes are linked to their primary outcomes. This includes the scope of the review (i.e., definition of research questions), conducting the search using the PICOC methodology [25] (i.e., queries definition), the bibliometric analysis, screening papers, and the classification scheme (i.e., taxonomy). The bibliometric analysis is not the core of the methodology, but it is used to ground a better understanding of the topic trends. It is developed in the following subsections as shown in the figure.

## 3.1 Search Methodology

To assess the status of metrics and KPIs for data monetisation, a systematic approach was used. The process starts with the definition of the key pillars that drive the topic and the use of targeted keywords to conduct literature searches [115, 116].

To ensure the relevance of our queries, we used thesaurus tools to identify terms closely related or similar to *dataset*, *value*, and *data centres*. We then used the IEEE Xplore search engine to search for these terms and identify the most commonly used. Results are presented in Table 3.

Table 3. Most relevant terms and their frequency grouped by dataset, value, and stakeholder themes in the context of data monetisation. (These searches were conducted in June 2024.)

| Dataset | | Value | | Stakeholder | |
| --- | --- | --- | --- | --- | --- |
| **Term** | **Freq.** | **Term** | **Freq.** | **Term** | **Freq.** |
| Dataset | 328,996 | Value | 303,174 | Information System | 45,172 |
| Data-set | 137,134 | Cost | 201,684 | Data warehouse | 5,304 |
| Data set | 76,056 | Price | 26,816 | Data platform | 1,904 |
| Data collection | 30,279 | Market price | 12,218 | Data centre | 916 |
| Repository | 13,321 | Commercial value | 5,679 | Data lakes | 152 |
| Data warehouse | 5,304 | Worth | 5,231 | — | — |
| Data record | 845 | Valuation | 1,280 | — | — |
| Data catalogue | 267 | Monetization | 198 | — | — |
| Data banks | 100 | — | — | — | — |

For the *dataset* theme, we retained all major keywords except for "Data record," "Data catalogue," and "Data banks," as these terms were less frequently mentioned and did not strongly align with the primary focus on data monetisation. The most significant terms, like "Dataset," "Data set," "Data-sets," "Data collection," "Repository," and "Data warehouse," showed high frequencies (e.g., "Dataset" appeared 328,996 times), indicating their strong relevance in the context of data assets and storage management.

In the *value* theme, we excluded the term "Price" from our analysis. Although "Price" appeared frequently (26,816 occurrences). It was observed that this term often led to topics outside the scope of metrics and KPIs related to data monetisation. To maintain a focused analysis, we prioritised terms like "Value," "Cost," "Worth," "Valuation," "Market Price," "Commercial Value," and "Monetization," which more directly relate to financial and economic perspectives on data.

This selective approach ensured that the retained keywords were highly relevant to data monetisation metrics and KPIs, avoiding unrelated topics and improving the accuracy of initial query results.

These keywords were used as markers, enabling efficient search across various platforms. Keyword selection was approached by establishing specific groups to help with the definition of the components of the keywords, taking a bottom-up approach. The methodology applied to construct queries and the definition of keywords follows the PICOC methodology – Population, Intervention, Comparison, Outcomes, and Context [25]. Each of these topics is defined as follows:

- *Population:* It refers to the specific group of individuals or subjects under the interest of the study. In the context of this study, we used the following keywords for representing the population of stakeholders that can trade and maintain (not necessarily use) data for different goals - data centre; data platform; data lakes; data warehouse; information system.
- *Intervention:* The intervention refers to the approach or technique applied in the empirical study. Here, the manipulation is the intrinsic KPIs and metrics over the data, and their calculation is the intervention process, leading to the use of metric, KPI, key performance indicator, and dimension. We also consider their use and classification, which led to including framework, model, system, and theory.
- *Comparison:* The comparison component involves differentiating methods, processes, or strategies. Since KPIs are closely related to strategies, KPI families could be further detailed into clusters. Nevertheless, since there is no clear classification or relation between KPIs, metrics, and strategies for data monetisation at this point, no further considerations were done for this component.
- *Outcomes:* Since empirical approaches or comparisons are not considered, no specific outcomes are defined in this component.
- *Context:* In our study, we have considered two types of context: the data itself and its value. We have thus separated the context into two subcategories of keywords, first in the theme of data with dataset and similar words, and second in the theme of data value with data valuation, data asset, information value, knowledge asset, and monetisation.

By combining the results from the PICOC approach, and specific keywords derived from systematic surveys on data value (e.g., [13]) the different keywords for the initial queries were built. The keywords families used in this work are described in Table 4.

Table 4. Keywords used for the search queries. We use two contexts to refine further our queries.

| Root | Keywords |
|---|---|
| Population | data centre; data center; data platform; data lakes; data warehouse; information system |
| Intervention | model; framework; system; theory; dimension; metric; KPI; key performance indicator |
| Comparison | assess; assessment; assessing; measure; measurement; measuring; evaluate; evaluation; evaluating; estimate; estimation; estimating |
| Outcomes | — |
| Context-data | dataset; data-set; data set; data collection; repository; data warehouse |
| Context-value | data value; data valuation; infonomics; data asset; value of data; data governance; information value; value of information; information valuation; knowledge asset; monetization; monetisation; data governance |

## 3.2 Bibliometric Analysis

For our study, we decided to focus on three main engines: Scopus since it seems to include the largest collection of articles, and ACM Digital Library and IEEE Xplore for their focus on computer science. Table 6 presents the findings before the screening phase, structured on publications per year when performing a broad analysis. These findings were obtained through the combination of queries on Population, Intervention, Comparison, Context-value, and Context-data. The main query used on each search engine is provided in listing 1.

Listing 1. Search query used to retrieve relevant literature.

```
("data centre" OR "data center" OR "data platform" OR "data lakes" OR "data
    warehouse" OR "information system") AND
("model" OR "framework" OR "system" OR "theory" OR "dimension" OR "metric"
    OR "KPI" OR "key performance indicator") AND
("assess" OR "assessment" OR "assessing" OR "measure" OR "measurement" OR "
    measuring" OR "evaluate" OR "evaluation" OR "evaluating" OR "estimate" OR "
    estimation" OR "estimating") AND
("dataset" OR "data-set" OR "data set" OR "data collection" OR "repository"
    OR "data warehouse") AND
```

```
("data value" OR "data valuation" OR "infonomics" OR "data asset" OR "value of
    data" OR "data governance" OR "information value" OR "value of information"
    OR "information valuation" OR "knowledge asset" OR "monetization" OR "
    monetisation" OR "data governance")
```

Further smaller queries were performed on the IEEE Xplore and the ACM digital library engines to extract more results. This was done using a lower number of keywords and sometimes adding some exclusion criteria such as NOT AI NOT "machine learning" in order to yield results more specific to our subject. Table 6 does not contain the results of these queries. The total number of results were 752 for IEEE and 2461 for ACM. The detail of these queries can be found in **??**.

Table 6. Number of publications found on IEEE Xplore, Scopus, and ACM Digital Library per year when using query 1. Searches where conducted between June 2024 and June 2025.

| Engine | 2014 | 2015 | 2016 | 2017 | 2018 | 2019 | 2020 | 2021 | 2022 | 2023 | 2024 | 2025 | Total |
|---|---|---|---|---|---|---|---|---|---|---|---|---|---|
| IEEE | 0 | 3 | 3 | 2 | 5 | 6 | 3 | 9 | 11 | 7 | 13 | 3 | 65 |
| Scopus | 202 | 214 | 230 | 248 | 317 | 348 | 490 | 585 | 716 | 895 | 1188 | 659 | 6092 |
| ACM | 117 | 119 | 133 | 134 | 117 | 174 | 206 | 290 | 516 | 478 | 606 | 216 | 3106 |

The search results show a steady increase in the popularity of the subjects covered in our study with more and more papers being published every year since 2014. Surprisingly, even though our keywords seemed to be relevant when using IEEE Xplore, this engine yielded a very low number of results when performing our queries. Both ACM Digital Library and Scopus provided a large number of results. Of course, there was some overlap in the search results between the engines and duplicates were filtered out during the screening process.

### 3.3 Screening Papers

A systematic analysis was conducted on the collected papers with a primary focus on assessing the relevance of the titles to the key facets identified previously. In cases where uncertainty arose, a more detailed analysis was performed based on the information provided in the abstract and the conclusion. This evaluation process followed a structured pipeline approach, as illustrated in Figure 3.

The rigorous screening ensured that each manuscript included at least topics connected to data valuation and monetisation or topics related to metrics and KPIs. Manuscripts that were not directly connected within data valuation and monetisation were also sometimes considered if metrics and KPIs could be relevant to the data environment.

### 3.4 Evaluation Results

From our evaluation of the literature retrieved, we retained 162 papers that fit our research questions. Tables 8-12 provide the list of metrics and KPIs we found. We have grouped under "Similar metrics" (second column) all the metrics that have a similar definition, synonyms, or metrics that represent an opposite (e.g., Objectivity vs Bias) and could thus be computed from one another (e.g., as $1 - M_{opposite}$ or as $1/M_{opposite}$ ). We only offer short definitions for the metrics for brevity's sake. The interested reader can obviously redirect themself to the cited references for a better description and grasp of the metrics listed here.

Answering RQ1 and RQ2, we observe that some metrics are much more common than others. In particular, metrics associated with data quality, e.g., accuracy (mentioned 43 times), completeness (mentioned 41 times), consistency (mentioned 36 times), or volume (mentioned 22 times) are at the centre of attention of many papers. High-quality data is essential for dependable insights and sound decision-making, serving as the backbone of successful data monetisation efforts, especially in data markets [44]. By ensuring that data meets high standards of quality, organisations can generate more accurate analyses and insights, impacting the reliability of monetisation/valuation outcomes. This is reflected in the numerous data quality ISO standards (e.g., ISO 8000-1:2022) and data quality frameworks that exist [104, 105, 132].

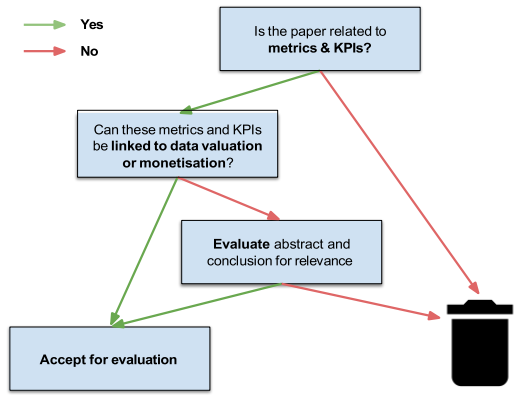

Fig. 3. Paper screening process.

Data timeliness, i.e. the availability of data when needed ensuring data is up-to-date and accessible within an appropriate time window, is also a critical point heavily discussed in the literature (mentioned in 42 references). Timeliness is a measurement of the delay between an event occurring and the data being available to the business, therefore it depends on age (mentioned 6 times). On the other hand, data currency (mentioned 37 times) is whether the data has lost its value due to its processing, modification or elapsed time and thus can be linked to change in data value with respect to time change or an event. Both timeliness and currency reflect how important it is for organisations to have access to data as soon as possible.

Metrics related to the ease with which data can be used, e.g. its usability, are also popular: accessibility (mentioned 20 times), clarity (23), conciseness (11), format (16), metadata (11), usability (14), and visualisation (4). This reflects the importance of documentation, clear interfaces and visualisations, and the overall understandability of datasets. Usability can be defined as a combination of several components. In fact, usability, according to ISO 9241: 11 (2018), is "a benchmarking tool that can be used to determine the extent to which a system, product, or service can be used by specific users to achieve the goals determined by the *effectiveness, efficiency, and satisfaction* of its users." [62]. Usability ensures that data not only meets technical standards (w.r.t. data quality and governance) but also aligns with the practical needs of users, enabling them to extract meaningful insights, make informed decisions, and complete tasks efficiently [36, 159].

Finally, metrics related to the overall trustworthiness of the data seem to also gain traction in the recent literature: integrity (16), lineage (5), objectivity (11), plausibility (21), reputation (17), traceability (14), and trustworthiness (12). Trustworthiness is used to evaluate the reliability and credibility of entities within a digital ecosystem. This metric is often derived from factors such as identity verification, transaction history, reputation, and credibility as perceived by transacting parties. It is a key metric in decentralised systems [20, 41, 51, 52, 153, 159] and in data markets [44].

## 4 A NEW TAXONOMY FOR KPIS AND METRICS

Given the high numbers of metrics found in the literature, and the relationship that exist between all of them, we decided to group them into a taxonomy. In this section, we present our use of the Balanced Scorecard (BSC) with further subclustering as a method of organising all the metrics we found into a single coherent taxonomy. The objective of this new taxonomy is to provide a new vision on KPIs and metrics that encompasses all the aspects of an organisation's business.

While we have tried to be as exhaustive as possible, we also know that finding all the metrics and KPIs defined in the literature is a gargantuan task. As such, our objective here is to help the interested reader navigate the vast landscape of metrics and KPIs that can be used in data valuation and monetisation, and continue the conversation on how these metrics can be organised to ease their understanding and usage in a business setting. We provide relationships between KPIs and metrics in this taxonomy, they might be subject to change from one organisation to the other. We have designed the relationships based on our literature review, and expertise from industrial partners. Finally, not every organisation will need (or be capable of measuring) every metric presented here. The idea here is to cover as much ground as possible and to leave to an organisation to select the metrics and personalise the taxonomy to fit their needs and use cases.

### 4.1 Balanced Scorecard (BSC)

The Balanced Scorecard (BSC) is a practical framework that helps companies to connect their day-to-day activities with their broader strategic goals [76]. The BSC, developed in 1990s, was meant to go beyond traditional financial performance metrics by including non-financial indicators, allowing organisations to track a more complete picture of their performance. Over time, it has become a powerful tool for aligning an organisation's long-term vision with clear, actionable goals. Having this ability to link a company's overall strategy to specific, measurable outcomes has made the BSC essential to many businesses aiming to achieve their strategic objectives.

Beyond aligning operations with strategy, the BSC is also highly effective for grouping and organising KPIs (or metrics). The BSC's structure of four main perspectives — Financial, Customer, Internal Processes, and Learning and Growth — categorises KPIs based on what part of the business they impact, making it easy to identify which KPIs are needed in each area. The following subsections present how each area contributes to aligning strategy with measurable outcomes.

The BSC is a flexible tool that helps organisations set up and track strategies focused on data monetisation. By using the BSC's four perspectives, companies can create a strategy that captures the value of their data assets. Breaking down each perspective into smaller, specific clusters further simplifies the process, making it easier to organise metrics that translate smoothly into KPIs. This approach not only keeps data monetisation efforts aligned with strategy but also makes tracking progress and outcomes more straightforward and consistent across different areas.

To facilitate tracking strategic achievements, we have defined different clusters. Metrics and KPIs are categorised based on the analysis performed on the different metrics and KPIs found in the literature. Figure 4 illustrates the higher levels of our taxonomy and how we divided the four BSC perspectives into clusters and subclusters. The taxonomy is composed of three layers: the higher-level data for executives and data owners managing an organisation's strategy, the middle-level data for data stewards handling the strategy at a department level for instance, and the low-level data for day-to-day operations and data specialists. The complete taxonomy is provided in Appendix ?? and will be fully detailed in future work, in particular on how all of these clusters interact with each other. The clusters and subclusters can be used as higher level KPIs that will be a weighted average on the metrics and KPIs it depends on. The weights can be computed based on preferences (e.g., in alignment with a specific strategy), or importance (e.g., how relevant each metric/KPI to the business).

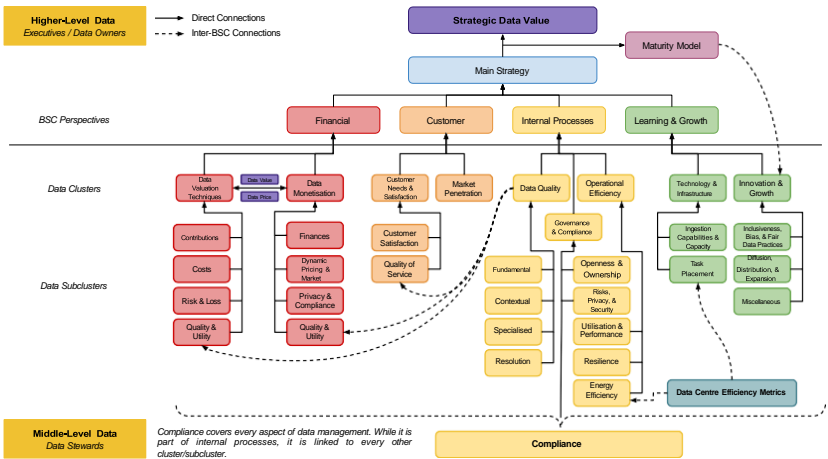

Fig. 4. The higher levels of our taxonomy. Each perspective is further divided into clusters and subclusters that will be used to categorise metrics and KPIs. Links between clusters are shown in dashed lines. The BSC perspectives and and the clusters can be used as KPIs based on a weighted average of the KPIs they depend on. Data centre metrics are numerous and specific enough that we decided to include them in their own subcluster.

In the taxonomy figures, BSC perspectives are colour coded to facilitate the identification of KPIs and metrics' category. Some metrics may appear in multiple clusters because they span multiple perspectives but they are associated to the colour of their main perspective. The following subsections present the clusters in more detail.

## 4.2 Maturity Model

A maturity model is a structured framework designed to assess the current state of data governance within an organisation. It identifies existing gaps and outlines the necessary steps to achieve a more advanced governance level. By aligning governance practices with strategic goals and evaluating their effectiveness, the maturity model ensures data management evolves to meet organisational needs. This model is typically tailored to specific domains, industries, and data types, enabling it to adapt to Big Data challenges [101].

Metrics play a critical role in feeding and informing the maturity model. Accurate and well-defined metrics provide the data-driven insights to evaluate progress and determine areas requiring improvement. Also, these metrics allow organisations to track their adherence to governance principles, ensuring alignment with compliance requirements and organisational goals. By doing so, they directly contribute to the estimation and refinement of the maturity model. For instance, the Capability Maturity Model (CMM) can indirectly be used as a valuation technique. A CMM enhances data value through a maturity-based framework that measures and improves data processes along dimensions like cost, quality, and utility [20, 52].

## 4.3 Financial Perspective

This perspective focuses on how a company's strategic choices impact financial outcomes. Metrics such as revenue growth, profit margins, and Return On Investment (ROI) help assess whether strategic actions are driving the expected financial benefits. By linking financial targets with strategic priorities, organisations can gauge their ability to deliver shareholder value, offering executives a way to evaluate how strategic shifts are influencing overall financial health. Table 8 presents the metrics and KPIs we categorised in this perspective.

Table 8. KPIs and metrics for the financial perspective.

| Metric / KPI | Similar metrics | Short explanation | References |
|---|---|---|---|
| Access Cost | Bandwidth Cost | The cost (not necessarily the price) incurred for accessing a dataset. | [125, 137] |
| Budget | Costs Planned, IT Plan | A budget outlines projected income and expenditures over a defined period to guide financial planning and goal achievement. | [58] |
| CAPEX | Capital Cost, Hardware Costs, Software/Application Costs, Service Cost, Infrastructure Unit Costs | Capital expenditure covers one-time, upfront investments in acquiring or upgrading physical assets and infrastructure, such as machinery, buildings, or for data - capturing hardware, acquisition systems, storage and processing equipment, valued at purchase cost or depreciation, with pre-installed software costs excluded. | [4, 13, 20, 68, 145, 168] |
| Data Acquisition Cost (DAC) | Cost of Procurement, Data Generation Cost, Cost of Collecting Data | The cost associated with the acquisition of data, whether through direct generation or data providers. | [16, 87, 121] |
| Data Price | Data Monetary Value, Price Function, Payoff, Reimbursement, Financial Value, Price of Information, Data Object Marketprice | The financial value assigned to data through market or model mechanisms; covering cost, payoff, reimbursement, and volume-based pricing. | [3, 16, 40, 63, 89, 93, 118, 121, 122, 150, 167–170] |
| Data Type | Data Usage Context, Data Scope | Classifies data into four types: A (operational), B (one-time decisions), C (legal/safety), and D (research/innovation). | [16, 145] |
| Data Value (DV) | Data Criticality, Value of Information, Fixed Record Value, IP Value, Intrinsic Record Value, Data Utility Value, Value of Data | Context-dependent relevance of data to a specific user or decision, while data criticality measures how indispensable that data is to operations and decision-making. | [6, 13, 16, 20, 26, 55, 63, 74, 78, 103, 145, 165, 169] |
| Data Value Ratio (VR) | — | The Data Value Ratio (VR) represents the proportion of a decision node's value attributed to a specific data source. | [145] |
| Demand | Score, Number of Data Consumers | Buyers' willingness to pay for data, quantified by desirability scores. | [20, 122] |
| Ease of Measurement | — | Readiness of a data value dimension to be quantified using existing methods and tools. | [21] |
| Economic Efficiency | Data Business, Characteristic Index | Weights data's operational relevance to gauge its criticality for business processes. | [30, 64] |
| Field Value | — | Relative importance of a dataset attribute to a business, model, or system. | [13] |
| Internal Rate of Return (IRR) | — | Rate that makes the net present value of cash flows zero. | [4] |
| Location YardStick Score | — | Contribution of location data to specific analytical tasks. | [75] |
| Loss and Missed Opportunity Costs | Data root cause remediation | Revenue and profit lost due to poor data quality. | [12, 165] |
| Loss of Information Value (LIV) | Revenue Loss, Replacement Cost, Lost Data Cost | Total financial impact of losing or compromising information. | [13, 16, 21, 55, 58, 121] |
| Market Adjustment Factor | Discount Price, Full Price | Market-condition adjustment factor applied to base data price. | [63, 168] |
| Market Value of Information (MVI) | Average financial contribution per record | Projected income from selling, leasing, or sharing information. | [20, 21, 52, 55] |
| Net Present Value | — | Present-value-based valuation of future cash flows. | [4, 145] |
| Node Value (NV) | — | Value contribution of a decision node weighted by data quality. | [145] |
| Open Data Barometer (ODB) | — | Global index measuring government open-data readiness, implementation, and impact. | [5] |
| Operational Cost (OPEX) | Maintenance Cost, System Cost, Storage Cost, Contractual Costs, Labor Costs, Utility Costs, Transaction Fees, Publishing Cost, Service Cost, Application Cost, Cost | Recurring costs of data-related operations (storage, maintenance, security, labor, utilities). | [7, 13, 26, 52, 58, 103, 121, 124, 137, 145] |
| Payment-Accuracy Tradeoff | — | Trade-off between payment for private data and accuracy of insights. | [163] |
| Privacy Budget | — | Privacy budget (epsilon) controlling trade-off between privacy and accuracy. | [57, 150] |
| Privacy Level | Privacy Sensitivity, Propensity Score | Privacy risk measured by how closely synthetic data resembles real data. | [49, 93, 150, 160, 163] |

*Continued on next page*

Table 8 – continued from previous page

| Metric / KPI | Similar metrics | Short explanation | References |
|---|---|---|---|
| Process Failure Costs | — | Frequency of process failures caused by poor data quality. | [12] |
| Processing Value Ratio | — | Value added by processing data into information. | [145] |
| Protection Expense | — | Cost to implement and maintain data protection measures. | [88] |
| Reconstruction Cost | Packet Recovery Score | Refers to the financial or computational effort needed to restore lost or corrupted data, typically from backups or redundant sources | [74, 137] |
| Revenue | Economic Benefits | Total income from normal business operations. | [3, 16, 51, 68] |
| Risk Cost | Regulatory Risk, Legislative Risk | Financial impact of a data breach, including fines and losses. | [16, 21, 86, 88, 121] |
| Scarcity | Data Rarity | Availability or rarity of information for monetisation. | [13, 16, 36, 121, 159, 168] |
| Shapley Value | Leave-One-Out (LOO), Feature Relevance, Feature Contribution | Marginal contribution of each data source in a coalition or feature in a machine learning process. | [8, 9, 12, 55, 60, 91, 96, 108, 164, 167, 168] |
| Space Cost | Disk Occupation | Storage space required per unit of data. | [69, 121, 161] |
| Storage Cost | Cost of data Storage | Infrastructure and management costs of data storage systems. | [16, 103, 137] |
| The Value of Information for Business (VIB) | Data Business Value, Applicability to Business | Actionable information quality (accuracy, relevance, timeliness). | [16, 55, 121] |
| Utility | Relevance, Application Characteristic Index, Retention | Functional contribution of data to achieving intended goals. | [11, 21, 26, 30, 42, 86, 105, 122, 145, 159, 173] |
| Value Added | Value-added, Feature Importance | Additional value added by a data source, a data point, a feature, or a column. | [12, 22, 58, 104, 105, 121, 126, 132, 135, 162, 167] |
| Value of Privacy | Privacy cost | Minimum payment required for specific privacy level. | [78, 150, 163] |
| Value Range | — | Potential value range achievable with required information. | [145] |
| Weighted Coverage Function | — | Price-adjustment weights for database query pricing. | [38] |

*4.3.1 Data Valuation Techniques.* Techniques, and its derived metrics and KPIs, that help assess the worth of data assets. They can help determine potential revenue-generating opportunities, calculate cost savings, define the relative value of data, and understand the economic impact of it. Given their relative impact, they can help define KPIs and metrics for data monetisation (next cluster). However, this does not necessarily directly transcribe into monetary gain because value differs from price.

Data valuation techniques include for instance decision-based valuation [145], using a CMM as a valuation technique [20, 52], geo-distributed data market valuation [125], or considering data as intangible assets [51].

*4.3.2 Data Monetisation.* KPIs and metrics that provide measurable evidence of the financial benefits derived from data initiatives. These metrics track the revenue generated through data-driven products or services, calculate cost reductions from improved decision-making, and gauge the ROI from data-focused projects.

The main metric, data price, can be calculated using various methodologies tailored to specific contexts and applications (e.g. [3, 89, 103, 118]). For example, entropy-based approaches provide a way to estimate the inherent value of data by measuring the reduction in uncertainty that the data contributes to a given model [89]. Elastic pricing models, as proposed in [3, 118], adapt prices dynamically based on supply, demand, and utility. These models ensure that pricing remains responsive to changing market conditions, providing a flexible framework for data valuation in competitive environments. Additionally, practical frameworks for data monetisation, like the one discussed in [103], integrate these theoretical methodologies into operational systems. These frameworks not only calculate data prices but also facilitate incorporation of implementation challenges, such as user incentives, privacy, and computational efficiency. This ensures applicability in real-world scenarios.

## 4.4 Customer Perspective

Recognising that customer satisfaction and loyalty are foundational for long-term success, the BSC emphasises metrics like customer satisfaction scores, market share, and brand loyalty. These indicators help track how effectively the company meets customer needs and provides value. By linking customer-oriented objectives with broader strategy, the BSC helps organisations focus on building lasting customer relationships. Table 9 provides the list of metrics we categorised under this perspective.

Table 9. KPIs and metrics for the customer perspective.

| Metric / KPI | Similar metrics | Short explanation | References |
|---|---|---|---|
| Access Frequency | Number of Requests, Arrival Rate, Access Interval, Number of Accesses, Usage Over Time, Data Usage Frequency | Total count of operations or queries made to a system within a given period. | [13, 16, 20, 30, 52, 66, 119, 145, 147] |
| Churn | — | Refers to the loss of customers. Prediction uses demographics, usage, transactions, interactions, sentiment and external factors to anticipate customer loss. | [67] |
| Competitive Advantage | Data Disclosure Impact, Exclusivity, Rival Access Loss | The degree to which data provides a unique strategic advantage, measured by the potential consequences if competitors gain access to it or if it is lost. | [16, 21, 86, 121] |
| Downloads | Download Frequency | The number of times users have clicked to retrieve a dataset. | [20, 34, 52, 119] |
| Learnability | — | Ease with which new users learn to use platform features like data search and visualisation. | [111] |
| Quality of Service (QoS) | Data Connected to Service Levels, Service Characteristic Index, Service Level Agreement | Performance level and reliability as defined by SLAs. | [30, 137, 165] |
| Reputation | Popularity, Authority | Perceived quality, governance, transparency, and ethical use of data. | [6, 12, 22, 44, 46, 73, 104, 105, 121, 126, 131, 135, 140, 146, 152, 159, 162] |
| Satisfaction | Feedback, User's Satisfaction, User Attitude, Behaviour, Business User Satisfaction, Degree of Satisfaction, Business User Satisfaction | User satisfaction with interface, layout, and data presentation. | [20, 34, 49, 52, 62, 70, 99, 111, 112, 138, 153, 174] |
| User Frequency | User Count, Concurrent Users, Number of Data Users, Data User Count | Concurrent user count and associated incremental cost. | [13, 16, 40] |
| Views | — | Number of times a dataset's page is viewed. | [52, 119] |
| Visualisation | — | Effectiveness of graphical data representations. | [34, 85, 105, 138] |
| Winning Rate | — | Winning rate of data providers in query allocation. | [58] |

*4.4.1 Customer Needs and Satisfaction.* Evaluates the alignment of data products or insights with customer needs and expectations. These indicators include customer retention, engagement, and/or satisfaction tied to data-related services. For instance, churn, referring to the discontinuation or cancellation of a service by a customer, can be measured and/or predicted to estimate a potential loss in revenue [67].

User frequency, concurrent users, and users number measure the regularity with which individual users engage with a platform or service within a specific time frame. These metrics reflect individual behaviour patterns, such as daily, weekly, or monthly activity, and are crucial to gauging user engagement and loyalty [13, 40]. High user frequency indicates active and returning users, which is vital for monetisation models based on subscriptions, premium features, or targeted advertising.

*4.4.2 Market Penetration.* Indicators and metrics that assess the extent to which data solutions have penetrated target markets, measuring factors such as market share growth and the adoption rates of new data products. These KPIs allow companies to gauge the reach and impact of their data monetisation strategies within the market, ensuring that these initiatives are effectively attracting and retaining customers.

Metrics like downloads and views, obtained by measuring how frequently datasets (or part of them) are accessed or used can help organisations gauge initial user interest and ongoing relevance. High download and view counts suggest that a dataset holds considerable appeal and reaches a broad audience. For example, on public sector platforms, datasets such as transportation or environmental statistics demonstrate a high level of engagement, helping organisations prioritise these resources for maximum impact [20, 34, 52, 119]. Other metrics, such as reputation or popularity, can also be important for an organisation image and help drive the monetisation of their data.

## 4.5 Internal Processes Perspective

This area examines the efficiency and effectiveness of essential internal processes that impact customer experience and financial performance. Typical metrics might include product development timelines, production efficiency, or quality standards. By prioritising improvements in these internal processes, the BSC encourages companies to refine operations in ways that support strategic objectives. This alignment ensures that internal operations are structured to meet strategic goals, which ultimately fosters a more streamlined and competitive organisation. Table 10 provides the list of metrics and KPIs we decided to categorise into this perspective.

Table 10. KPIs and metrics for the internal processes perspective.

| Metric / KPI | Similar metrics | Short explanation | References |
|---|---|---|---|
| Accessibility | Search, Functionality, Navigation, Link | Ease with which users can access and utilise data. | [12, 16, 19, 22, 29, 34, 46, 86, 104, 105, 120, 125, 126, 135, 137, 146, 159, 162, 171, 175] |
| Accuracy | — | How closely data or results align with the true, intended, or expected values. | [3, 10–12, 14, 16, 17, 22, 23, 31, 34, 46, 49, 54, 55, 57, 59, 64, 73, 77, 86, 87, 90, 99, 102, 104, 105, 109, 111, 121, 126, 134, 135, 140, 145, 146, 148, 153, 159, 160, 162, 163, 174] |
| Adaptability | Versatility, Flexibility | Measures a system's ability to adjust to environmental changes, like network conditions or user demands, to enable dynamic resource allocation and efficient scalability. | [62, 104, 131, 144, 152, 153, 159] |
| Age | Data Creation, Time Index, Age of Information | Refers to how recent or old the data is. The age of data can impact its relevance and accuracy, especially in fast-moving fields like technology or finance. Related to Timeliness. | [6, 29, 51, 63, 121, 132] |
| Availability | Retrievability | Availability measures how reliably authorised users can access data when needed, encompassing system reliability, fault tolerance, backups, and redundancy. | [16, 34, 55, 68, 73, 81, 95, 104, 105, 112, 120, 121, 126, 135, 175] |
| Clarity | Understandability, Easy to Understand, Ease of Understanding, Lack of Confusion, Unambiguity, Readability, Interpretability, Usability | Data Clarity describes how organised, consistent, readable, and well-contextualised data is, enabling users to interpret and use it without confusion. | [5, 11, 12, 18, 22, 36, 46, 54, 55, 62, 80, 90, 104, 105, 111, 121, 126, 135, 140, 146, 162, 164, 171] |
| Class Overlap | Class Similarity, Euclidean Minimum Spanning Tree, Davies-Bouldin Index (DBI) | Measures how much classes or clusters in a dataset overlap based on regions of data points containing multiple classes or clusters (based on the label associated to the points). | [61, 128, 142] |
| Completeness | Appropriate Amount of data, Incompleteness, Sparsity Rate, Comprehensiveness | Gauges how fully a dataset contains required information - ensuring mandatory fields are populated, missing values are minimal, structure matches expectations, and granularity meets analysis needs. | [11, 12, 16, 17, 20, 22, 23, 29, 36, 46, 52–55, 58, 59, 64, 66, 77, 78, 80, 81, 86, 90, 102, 104, 105, 121, 126, 129, 131, 134, 135, 146, 148, 149, 153, 159, 162, 173, 175] |
| Conciseness | Concise Representation, Simplicity | Captures how briefly and clearly information is conveyed by minimising redundancy and noise while preserving essential content. | [12, 46, 55, 104, 105, 111, 121, 131, 135, 140, 146] |
| Confidence | — | Measures how strongly a dependency holds in the dataset. | [155] |

Table 10 – continued from previous page

| Metric / KPI | Similar metrics | Short explanation | References |
|---|---|---|---|
| Consistency | Linked to Heterogeneity, Veracity, Semantic Consistency, Syntactic Consistency, Representational Consistency, Coherence | How well data adheres to predefined rules. | [8, 10–12, 14, 17, 18, 22, 23, 36, 46, 53, 54, 59, 64, 66, 80, 86, 90, 94, 102, 104, 105, 121, 126, 129, 132, 134, 135, 146, 148, 164, 165, 171, 173, 175] |
| Containment Fraction | Dataset Overlap | How much of one dataset is encompassed by another. | [137] |
| Correctness | Label Purity, Label Noise, Noisy Labels, Free of Error | Measures the inconsistencies in labels (e.g., in annotated data), indicating how many data point have values corresponding to the real-world. | [12, 61, 105, 126, 132, 175] |
| Cost of Degradation (CoD) | − | Quantifies the loss of data quality due to privacy-preserving transformations. | [143] |
| Currency | Linked to Timeliness, Currentness, Up-To-Date, Diminishing Value, Decay | Gauges data freshness by comparing its age against its required update frequency | [5, 6, 11, 12, 19, 21, 22, 34, 36, 53, 55, 59, 63, 64, 66, 73, 77, 78, 90, 99, 102, 104, 105, 120, 126, 129, 132, 135, 140, 146, 148, 149, 153, 159, 162, 168, 171, 173] |
| Data Principles and Practices | Data Standard, Standardisation, Compliance | Measures compliance with defined data standards and governance policies to ensure data consistency, reliability, and strategic alignment. | [64, 104, 105, 135, 159, 165] |
| Data Similarity | Euclidean Distance, projection similarity, similarity score, cosine similarity, average distance, Kolmogorov-Smirnov, Mann-Whitney, Mood's Median, Levene's test (LE), $\chi^2$-test, negative log-likelihood | Distance metrics quantifying numerical distances between points rather than relying on syntactic likeness; for example, projection similarity assesses datasets by comparing their feature dimensions. | [6, 29, 35, 43, 85, 108, 155, 160] |
| Data Robustness | Shapley Robust | Ability of data to maintain value and usability across tasks. | [3] |
| Detail | — | Level of data detail - its granularity and precision. | [55] |
| Differential Privacy | — | Limits how much an algorithm's output can change when a single individual's data is added or removed. | [42, 93, 150, 163, 167] |
| Expected Annual Fraction of Data Loss (EAFDL) | Annual Fraction of Data Loss (AFDL), Mean Time to Data Loss (MTTDL) | EAFDL measures annual data loss as a fraction of total stored data based on the mean time to data loss, as an indicator of storage reliability. | [72] |
| Elapsed Time | | Total time taken to complete an operation. | [65, 71, 118] |
| Encryption Time (ET) | Decryption Time (DT) | Time required to encrypt or decrypt data. | [43] |
| Entropy | Shannon's Entropy, Heterogeneity, Information Entropy, Additional information Value (AIV), Joint Entropy, Individual Entropy, Information Score | Measures dataset randomness or information content. | [9, 13, 38, 66, 85, 89, 105, 149, 155, 164, 168–170] |
| Error Rate / Ratio / Count | Uplink / Downlink Error Rate, Trouble Tickets, Inter-Server Error Rate (ISER), Failure Rate | Proportion of operations that fail. | [16, 20, 50, 52, 57, 69, 73, 147, 157] |
| FAIRness Score | — | Evaluates datasets' compliance with the FAIR principles. | [92] |
| Format | Format Compliance, Codification, Conformity, Available Formats, Format Homogeneity, Format Validity, Value Data Type, Data Portability, Uniformity | Measures adherence to specified data formats, making the data usable across systems. | [17, 19, 34, 61, 80, 98, 99, 102, 104, 105, 121, 135, 138, 153, 159, 175] |
| Granularity | Data Frequency, Abundance, Time Resolution | Degree of detail or precision in data. | [34, 42, 94, 100, 105, 132, 155] |
| Inferential Privacy | — | Probability an adversary correctly infers a private parameter from observed data. | [42] |
| Information Content (IC) | − | Unique information content per data packet. | [74] |
| Information Frequency | — | Rate at which information is updated, accessed, or used. | [145] |
| Integrity | Reliability, Data Prevention, Data Source, Corroboration | Accuracy, consistency, and reliability of data across its lifecycle. | [7, 12, 34, 36, 55, 59, 64, 73, 95, 104, 131, 146, 159, 162, 165, 171] |
| Interoperability | Compatible, Integration Capabilities | Ability of systems to seamlessly share and utilise data across platforms. | [12, 18, 94] |
| Latency | Uplink/Downlink Communication Latency, Interserver Communication Latency, Database Access Latency, Transaction Finality Time, Link/Downlink Communication Latency (UDCL) | Time between a request and its response. | [6, 7, 27, 50, 65, 69, 74, 117, 118, 125, 134, 157] |

Table 10 – continued from previous page

| Metric / KPI | Similar metrics | Short explanation | References |
|---|---|---|---|
| Licensing | License Compliance, Free License, Licensing Restrictions, Usage Restrictions | Share of datasets published under specified licensing terms. | [5, 18, 19, 31, 34, 86, 98] |
| Lineage | Provenance | Complete record of a dataset's origin and all transformations. Refers to the what and how. | [26, 73, 98, 105] |
| Maintainability | — | Ease of managing, updating, and enhancing data systems. | [12, 153] |
| Maintenance Frequency | — | Frequency with which a dataset requires maintenance. | [137] |
| Metadata | Contextual Information, Documentation, Documentation Features, Profiling | Detailed specification of required information format and medium. | [18, 31, 44, 62, 77, 80, 98, 104, 121, 145, 148] |
| Moderation | Typicality, Outliers, Variability | Proportion of data points within the extremes of a normal distribution. | [61, 102, 105] |
| Mutual Information | — | Mutual information shared between two variables (from a probabilistic point of view) | [6, 13, 33, 85, 89] |
| Network Metrics | — | Operational-efficiency metrics for data-centre networks (e.g., stretch, size, utilisation). | [155] |
| Network Traffic Overhead | — | Bandwidth and resource consumption per data transaction. | [7] |
| Number of Sensitive Field | — | Count of governed data columns in a data store. | [88] |
| Objectivity | Bias, Unbiasedness, Fairness, Free from Bias | Degree to which data or a data source is believed to be free from biases and impartial. | [12, 15, 22, 31, 36, 61, 104, 105, 121, 135, 162, 175] |
| Openness | Sharing | Degree to which datasets are openly licensed and accessible. | [20, 81, 90, 120, 136, 159] |
| Oversight | Audit | Capability to monitor and audit compliance with governance policies. | [165] |
| Ownership | — | Ownership clarity and licensing restrictions of a dataset. | [34, 58, 121] |
| Plausibility | Credibility, Believability, Match Between the System and the Real World | Degree to which data is credible and aligns with expected patterns or the real world. | [12, 22, 36, 46, 54, 55, 62, 77, 104, 105, 121, 132, 135, 140, 146, 149, 159, 162, 171, 173, 174] |
| Policy | — | High-level principles guiding organisational decision-making and behaviour. | [88, 165] |
| Precision | — | Level of detail with which data is captured and represented. | [12, 36, 104, 105, 132, 135, 145, 146, 153] |
| Proximity | — | Proximity relevance of data based on event or sensor location. | [6] |
| Purity | — | Measures the importance of a dataset in a network of datasets based on its centrality and quality. | [17] |
| Quality Factor (QF) | — | Link between information quality and its business innovation potential. | [145] |
| Range | — | Quantifies the proportion of data values that fall within predefined lower and upper bounds, reflecting the validity of data within expected limits. | [64, 102, 105] |
| Recoverability | Backup, System Backup, Recovery Capabilities | Assesses a system's secure data preservation and rapid recovery capabilities during disruptions. | [68, 104, 105, 135] |
| Regulatory Compliance | Compliance Cost, Legal Compliance | Degree of legal and regulatory compliance of data. | [57, 79, 131, 140, 146, 171] |
| Relevance | Relevancy, Decision Support Capabilities, Relevance Factor, Priority Score, Importance, Existence | Usefulness and adaptability of data for business processes. | [6, 12, 13, 16, 22, 30, 36, 46, 54, 55, 68, 73, 78, 94, 104, 126, 135, 145, 146, 148, 153, 162, 173] |
| Response Time | Latency | Total latency from request initiation to system response. | [16, 68, 71, 95, 118, 121, 134, 154] |
| Responsiveness | Time Metrics, Speed | Responsiveness of a system to inputs beyond raw processing speed. | [12, 14, 55, 62, 68, 69, 153, 174] |
| Risk Score | Risk Management Index | Overall risk score of a data store based on multiple risk factors. | [88, 94] |
| Runtime | Processing Time | Runtime overhead relative to a performance baseline. | [134, 144] |
| Scalability | System Concurrence Processing Capabilities, Elasticity | System's ability to scale resources dynamically with demand. | [35, 38, 68, 95, 117, 137, 144, 152] |
| Schema | — | Presence of desired data attributes like clarity and robustness. | [12, 24, 35] |
| Security Composite Efficiency Indicator | — | Composite cybersecurity efficiency score (readiness, resilience, penetration). | [79] |
| Security | Security Level Index, Access Security, Access Control, Data Protection, Confidentiality | Strength of data protection measures (confidentiality, integrity, availability). | [12, 22, 30, 34, 44, 46, 68, 78, 79, 104, 105, 121, 126, 135, 171] |
| Service Agreement | — | Clarity and enforceability of data - usage contracts and licenses. | [58, 121] |

Table 10 – continued from previous page

| Metric / KPI | Similar metrics | Short explanation | References |
| --- | --- | --- | --- |
| Shapley Fairness | Fairness Metrics | Fair allocation of resources or data access among users or groups. | [3, 71, 92] |
| Stochastic Divergence | Identity-based Exact Match, Jensen-Shannon Divergence, Wasserstein Distance, Hellinger-distance, Confidence Interval Overlap, KL-Divergence | Measure of distribution similarity or divergence between datasets. | [35, 85, 103] |
| Structure | Data Structure | Organisation and format of stored data for efficient access. | [12, 20, 35, 159] |
| Success Rate / Ratio / Count | — | Operational success rate: completed tasks versus attempts. | [87, 147] |
| Support | — | Support for database rule mining: proportion of records matching rules. | [155] |
| Synchronisation | — | Degree of synchronisation between multiple data sources. | [2] |
| Syntactic Similarity | Levenshtein Distance, Edit Distance, Cosine Similarity, Q-gram Distance, Semantic Similarity | String similarity metrics for data value comparison. | [155, 164] |
| System Robustness | System Stability | System stability and resilience under disturbances or attacks. | [68, 87] |
| System Utilisation | CPU Utilisation, Memory Utilisation, Disk Utilisation, Deployed Hardware Utilisation (DH-UR) | Utilisation rate of CPU, memory, storage, and network resources. | [45, 65, 69, 123, 147, 154] |
| Throughput | Transaction Processing Speed | Throughput: number of requests processed per time unit. | [6, 7, 65, 69, 78, 117, 118, 154] |
| Timeliness | Data Freshness, Punctuality | Data availability and timeliness for user needs. | [5, 6, 11, 12, 16, 19, 21–23, 29, 34, 36, 46, 53–55, 59, 63, 64, 66, 73, 77, 78, 86, 90, 99, 102, 104, 105, 121, 126, 129, 132, 135, 140, 146, 148, 149, 153, 159, 162, 173] |
| Traceability | Addressability, NGD, NDI, NDG, NID, NDGI, Verifiability, Provenance Documentation, Audit Trail Coverage, Compliance Rate, Watermarking | Ability to trace data lineage, provenance, and audit trails. Refers to the who, where, and when. | [12, 28, 31, 44, 73, 82, 98, 102, 105, 132, 135, 141, 162?] |
| Usability | Usage, Ease of Use, Easy to Use, Ease of Manipulation, Ease of Operation, User Friendliness, Ease of Search, Manipulability | Ease of accessing, understanding, and using quality data. | [12, 34, 36, 46, 58, 62, 68, 81, 99, 104, 144, 153, 159, 173] |
| Validity | — | Adherence of data to defined business rules, criteria, and standards. | [29, 36, 53, 54, 59, 64, 73, 104, 126, 171] |
| Variety | Multifacetedness | Diversity of data types and formats. | [14, 35, 58, 67, 132, 149, 171, 175] |
| Velocity | Frequency Parameter | Rate of data generation, ingestion, and processing. | [30, 66, 105, 145, 154, 171] |
| Volatility | — | Refers to how frequently data changes or how long it remains valid before becoming outdated. | [12, 105, 134] |
| Volume | Quantity, Entries, Dimension, Amount of Data, Size | Total volume of data available for analysis. | [9, 14, 16, 20, 22, 46, 52, 58, 67, 69, 73, 94, 98, 104, 105, 121, 135, 138, 140, 155, 162, 171] |

4.5.1 *Data Quality.* Indicators that measure the accuracy, completeness, consistency, and other quality-related measures/indicators of data. High-quality data is essential for dependable insights and sound decision-making, serving as the backbone of successful data monetisation efforts. By ensuring that data meets high standards of quality, organisations can generate more accurate analyses and insights, impacting the reliability of monetisation outcomes.

Data quality is a complex subject in itself and has been discussed and theorised extensively in the literature through ISO standards (e.g., ISO 8000-1:2022) and data quality frameworks [105, 132]. Survey [12] alone references 76 data quality dimensions. We have decided to subdivide data quality into 4 more subclusters:

- *Fundamental quality metrics* refer to metrics that can be used in any context such as age, metadata, completeness, granularity or precision.

- *Resolution quality metrics* refer to the granularity, the information frequency, or the timeliness and currency of the data.
- *Contextual quality metrics* refer to metrics whose definition depend on the context or the application, such as typicality, usability, clarity, or conciseness.
- *Specialised quality metrics* refer to specific data processes such as machine learning. This includes metrics such as similarity, class overlap, fairness, or stochastic divergence.

*4.5.2 Data Governance and Compliance.* Indicators and metrics that monitor adherence to data governance principles and regulatory standards, such as data privacy and security requirements. By aligning data practices with legal and ethical guidelines, these metrics reduce the risks associated with data breaches and regulatory non-compliance, thus fostering trust with clients and stakeholders.

Compliance involves metrics and KPIs designed to track adherence to legal, standards, and regulatory requirements such as regulatory compliance (e.g., ISO 20022), internal and external compliance, or oversight. As illustrated in Figure 4, compliance is ubiquitous and every other cluster or subcluster depends on it, covering every aspect of an organisation's business.

Data governance refers to the definition of processes, methods, roles, policies, standards, and metrics for managing and ensuring the proper use, discovery, collection, processing, analysis, disposal, and storage, of data within an organisation [101]. By implementing sound governance, data can be accurate, secure, consistent, and accessible for authorised use while meeting compliance requirements and supporting organisational goals.

*4.5.3 Operational Efficiency.* KPIs and metrics that evaluate how effectively data is integrated into the company's operational processes, assessing the impact of data on productivity and resource management. These indicators help track improvements in process efficiency due to data utilisation, ensuring that data-driven operations are streamlined and resources are optimised for maximum productivity. We have subdivided this cluster into utilisation & performance, resilience, and energy efficiency.

Utilisation & performance contains several key performance metrics in computing, data management, and networking which can be attributed to either technology & infrastructure or operational efficiency, depending on their context. Technology & infrastructure (discussed in a following section) focuses on the hardware, network, and system architecture that define the technical capabilities and limitations of a system. In contrast, operational efficiency refers to how effectively those resources are managed, tracked, optimised, and utilised to enhance performance, as demonstrated in a few real scenarios such as Customer Relationship Management (CRM) data analysis and high-speed railway data management [27, 43, 149, 154]. Given these considerations, metrics like latency, bandwidth, network traffic overhead, and throughput are linked to both clusters.

Resilience focuses on the reliability and the resilience of data systems under pressure and encompasses metrics such as system robustness, availability, error/success rate, maintainability, etc., while energy efficiency is linked to the efficient management of a data centre.

*4.5.4 Data Centre Efficiency Metrics.* Even though we would categorise these metrics into operational efficiency, their sheer number and specificity made us separate them into their own category. These metrics concern the management of a data centre in terms of power usage, hardware deployment and usage, and how efficiently a data centre is managed. Table 11 lists all the metrics we found in the literature on this subject. Surveys [123, 156] were key in the list of metrics found in this cluster and we encourage the interested reader to read these works. The most well-know and widely used metric is probably the power usage effectiveness (PUE), defined in ISO/IEC 30134-2 / EN 50600-4-2. We have included numerous derivatives that have been developed over the years.

Table 11. KPIs and metrics for data centre management and efficiency.

| Metric / KPI | Similar metrics | Short explanation | References |
|---|---|---|---|
| Adaptability Power Curve (APC) | Adaptation of Data Centre to Available Renewable Energy (APCren) | Deviation of a data centre's power use from its baseline | [114, 158] |
| Air Economiser Utilisation Factor (AEUF) | — | Assess how effectively a data centre employs its airside economiser system for "free" cooling. | [123, 133] |
| Airflow Efficiency (AFE) | — | How efficiently air moves from the supply to the return. | [151] |
| Carbon Emission Factor (CEF) | — | Coefficient used to calculate the amount of $CO_2$ emissions produced per unit of energy consumed. | [32] |
| Carbon Usage Effectiveness (CUE) | — | Measure the environmental impact by assessing the amount of $CO_2$ emissions per unit of IT energy consumed. | [32, 91, 114] |
| CO2 Savings | — | CO2 Savings measures the reduction in a data centre's carbon emissions relative to a baseline achieved through energy optimisations, renewable integration, and heat recovery. | [114, 123, 151] |
| Compute Power Efficiency (CPE) | — | Metric that assesses the efficiency of IT equipment in a data centre by measuring how much of the total power consumption is effectively used for computation. | [156] |
| Cooling Capacity Factor (CCF) | — | Assess the utilisation efficiency of the cooling infrastructure. | [151] |
| Cooling Effectiveness Rate (CER) | Energy Effectiveness of Cooling | Focuses on the effectiveness of cooling systems in a data centre. | [114] |
| Corporate Average Data Centre Efficiency (CADE) | — | Comprehensive assessment of a corporation's data centre energy performance by considering both infrastructure efficiency and IT asset utilisation. | [123, 133] |
| Data Centre Adaptation (DCA) | Data Centre Energy Profile Change | Measures the change in a data centre's energy profile from a baseline to assess its efficiency and adaptability. | [114] |
| Data Centre Compute Efficiency (DCcE) | — | Ratio of computational work performed by a data centre's IT equipment to the energy consumed | [123] |
| Data Centre Lighting Density (DCLD) | Lighting Power Density (LPD) | Is the lighting power consumption per unit area (expressed in W/ft2 or W/m2) to evaluate a data centre's lighting efficiency. | [123, 133] |
| Data Centre Performance Efficiency (DCPE) | — | Metric used to evaluate how effectively a data centre utilises energy to perform useful work. | [123] |
| Data Centre Performance Per Energy (DPPE) | Data Centre Energy Productivity (DCeP), Data Centre Productivity (DCP) | How efficiently a data centre converts energy into computational performance. | [123, 156] |
| Data Centre Power Density (DCPD) | — | Measure how much power is consumed per rack in a data centre. | [123] |
| Data Centre Space Efficiency (DCSE) | — | Physical space utilised per unit of available space. | [123] |
| Data Centre Workload Power Efficiency (DWPE) | Workload Power Efficiency (WPE) | IT equipment energy consumption divided by total data centre energy consumption. | [123, 166] |
| Δ-T Per Cabinet | T - Per Cabinet | Temperature difference between cabinet inlet and outlet air. | [151] |
| Deployed Hardware Utilisation Efficiency (DHUE) | — | Ratio of minimum servers required for peak load to total servers deployed. | [123] |
| Electronics Disposal Efficiency (EDE) | — | Assesses the disposal of decommissioned Information and Communication Technology (ICT) assets. | [123] |
| Energy Data Centre ($E_{DC}$) | — | Total energy consumption of a data centre, encompassing energy usage of both IT hardware and supporting infrastructure. | [133] |
| Energy Efficiency | Power to Performance Effectiveness (PPE), Power Usage Effectiveness (PUE), System Power Usage Effectiveness (sPUE), Data Centre Infrastructure Efficiency (DCiE), Data Centre Performance per Energy (DPPE), Data Centre Energy Productivity (DCEP), Communication Network Energy Efficiency (CNEE), Network Power Usage Effectiveness (NPUE), Energy Proportionality Coefficient (EPC) | Workloads processed per unit of energy consumed. | [32, 45, 66, 91, 114, 123, 130, 133, 156, 157, 166, 172] |
| Energy ExpenseS (EES) | — | Change in energy expenses relative to a baseline following upgrades or flexibility measures. | [123] |

*Continued on next page*

Table 11 – continued from previous page

| Metric / KPI | Similar metrics | Short explanation | References |
|---|---|---|---|
| Energy reuse factor (ERF) | Energy reuse effectiveness (ERE) | Defined in ISO/IEC 30134-6 / EN 50600-4-6, determines the share of the total energy consumption that is reused. | [32, 45, 66, 91, 114, 123, 130, 133, 156, 157, 172] |
| Energy Waste Ratio (EWR) | — | Proportion of energy in a data centre that does not directly contribute to IT operations. | [123] |
| Fixed to Variable Energy Ratio (FVER) | Data Centre Fixed to Variable Energy Radio (DC-FVER) | Assess the proportion of fixed energy consumption relative to variable energy consumption. | [123, 151] |
| Green Energy Coefficient (GEC) | — | Share of a data centre's total energy consumption sourced from renewables. | [123, 156] |
| Grid Utilisation Factor (GUF) | — | Proportion of power drawn from the electrical grid versus generated on-site. | [151] |
| HVAC Availability, Capacity, and Efficiency (ACE) | — | ACE in HVAC tracks: Availability, Capacity, and Efficiency via general mathematical expressions. | [123] |
| HVAC System Effectiveness | HSE | Overall efficiency of a data centre's cooling system. | [123, 151] |
| IT Equipment Energy Efficiency (ITEE) | — | Measures useful work performed by IT equipment per unit of energy consumed. | [123] |
| IT Equipment Utilisation (ITEU) | — | Ratio of actual to rated energy consumption of IT equipment. | [123] |
| IT Power Usage Effectiveness (ITUE) | — | PUE metric applied to IT equipment rather than the entire data centre. | [133] |
| Other HVAC Utilities | — | Efficiency of thermal and air management in data-centre operations. | [84, 91, 123, 151, 156, 157] |
| Power Density Efficiency (PDE) | — | Improvement in energy efficiency from rack-level physical changes. | [84, 123] |
| Power Usage Effectiveness (PUE) | Carbon Usage Effectiveness (CUE), Total Energy Usage Effectiveness (TUE), Data Centre Efficiency (DCE), Data Centre Infrastructure Efficiency (DCiE), SPUE, pPUE, PUE1-4, SI-POM, H-POM | Ratio of total facility power to IT equipment power (PUE) and related carbon and conversion metrics. | [32, 45, 91, 114, 156, 157] |
| Renewable Energy Factor (REF) | On-site Energy Fraction (OEF), On-site Energy Matching (OEM) | Share of energy consumption met by renewables, on-site and grid. | [154] |
| Server Compute Efficiency (ScE) | — | Compute efficiency of data-centre resources per energy consumed. | [123] |
| Space, Wattage, and Performance. (SWaP) | — | Overall facility resource usage including support infrastructure. | [32, 45, 66, 91, 114, 123, 130, 133, 156, 157, 172] |
| Stranded Power Capacity Per Rack (SPCR) | — | Unused power allocated within server racks. | [151] |
| The Green Index (TGI) | — | Flexible green benchmarking metric based on performance-per-watt. | [123] |
| Total Equipment Utilisation (TEU) | — | Overall resource usage of the entire facility. | [32, 45, 66, 91, 114, 123, 130, 133, 156, 157, 172] |
| Traffic Energy | Management and Monitoring Traffic Energy (MMTE) | Energy footprint of data collection, transmission, and processing. | [50] |
| Traffic Ratio | Management and Monitoring Traffic Ratio (MMTR), Internal Traffic, External Traffic | Proportion of specific traffic type in total network data. | [50] |
| Uninterruptible Power Supply (UPS) Metrics | UPS Surge Factor (USF), UPS Crest Factor (UCF), UPS Power Factor (UPF), UPS Power Factor Corrected (UPFC), UPS Power Energy Efficiency (UPEE) | UPS energy-conversion efficiency and related performance metrics. | [123, 151] |
| Water Usage Effectiveness (WUE) | — | Water usage per unit of IT energy consumption. | [123, 156] |

## 4.6 Learning and Growth Perspective

The BSC also highlights the importance of continuous development of organisational capabilities, from workforce skills to information systems and culture. Metrics in this area may involve employee training hours, knowledge-sharing practices, and skill enhancement programs. By connecting these learning-oriented objectives to strategic aims, the BSC underscores the importance of fostering an adaptable and skilled workforce, essential for sustaining innovation and a competitive position over time.

Table 12. KPIs and metrics for the learning and growth perspective.

| Metric / KPI | Similar metrics | Short explanation | References |
| --- | --- | --- | --- |
| Cache Size | — | Indicates the total volume of data a cache can hold (in KB, MB, or GB) to accelerate future data retrieval. | [71] |
| Data Ingestion Capabilities | Data Collection and Management Capabilities, Data Analysis Capabilities, Data Mining Capabilities | Describes a system's ability to ingest and store high-volume, low-latency data from diverse sources at scale. | [40, 68, 161] |
| Discoverability | Visibility, Findability, Platform Performance, Platform Effectiveness | Effectiveness of a platform in returning relevant datasets for user queries. | [62, 105, 111, 120] |
| Extensibility | — | Ability to add new features or devices without disrupting existing systems. | [18, 47, 152] |
| Growth Rate | — | Rate of increase in dataset record count over time. | [151] |
| Inclusiveness | — | How well an initiative leverages diverse stakeholder contributions while still achieving its goals. | [87] |
| Information Diffusion | Data Distribution, Information Distribution | Speed and scope of information sharing across entities. | [37, 159] |
| Lifecycle | Shelf life of data, Information lifetime, Period of Use, Data Longevity | Expected active lifespan of a data source. | [16, 20, 52, 145, 165] |
| Quantity of Public Projects | Quantity of Private Projects | Number of private vs. public projects/services/datasets available online. | [40] |
| Social Welfare | Social Benefits | Net utility gain minus privacy cost in a data marketplace. | [68, 150] |
| Statistical Parity | Class Parity, Class Imbalance | Fairness in machine learning via equal positive-class probability for sensitive groups. | [61, 109, 132, 175] |
| System Capacity | Memory, CPU, Bandwidth, Storage Capacity | Maximum load capacity before performance degradation. | [30, 32, 69, 91, 114, 123, 125, 134, 144, 154, 157] |
| Transparency | Objective Measurement | Clarity and interpretability of data and processes. | [80, 105, 109, 165] |
| Trustworthiness | Assurance, Trust, Trust Score | Trustworthiness of data based on reputation and credibility. | [20, 31, 41, 44, 52, 58, 105, 121, 126, 153, 159, 175] |

*4.6.1 Technology and Infrastructure.* KPIs and metrics that assess the organisation's capability to manage large-scale data initiatives, focusing on aspects such as data storage, processing power, and analytics infrastructure. This sub cluster ensures that the technical foundation for handling data monetisation is in place, enabling the organisation to store, analyse, and leverage data effectively. Investments in technology and infrastructure support scalability and enhance the organisation's capacity for data-driven innovation. As previously mentioned, several key performance metrics in computing, data management, and networking can be attributed to either technology and infrastructure or operational efficiency, depending on their context.

For technology and infrastructure, we identified two subclusters: ingestion capabability & capacity, and task placement. Task placement is directly related to data centre efficiency metrics and specific algorithms used in distribution, for example in the GENiC architecture [114]. Ingestion capabilities and capacity refer to a system's capacity to efficiently acquire, process, and store data. This includes metrics such as velocity, disk throughput, CPU performance, etc.

*4.6.2 Innovation and Growth-Oriented.* Indicators and metrics that measure the organisation's progress in fostering a culture of data-driven innovation, including the development of new data products, R&D spending on data initiatives, and employee engagement in data-focused projects. By tracking innovation-oriented KPIs, organisations can ensure they are continuously evolving to capitalise on emerging data opportunities, promoting long-term growth and adaptation in a competitive landscape.

This cluster KPIs and metrics that facilitate the measurement and tracking of system or enterprise growth, specifically by assessing the progression of different subsystems and their impact on overall business development. To provide a structured approach to tracking such evolution, maturity models play a crucial role in evaluating an enterprise's growth trajectory (hence the direct connection as shown in Figure 4). These models offer a staged framework that allows organisations to measure their progress across predefined levels — ranging from initial or ad-hoc implementation to optimised and data-driven maturity. By leveraging such models, enterprises can systematically monitor their progress, identify gaps, and strategically enhance their operational and analytical capabilities. This cluster also includes metrics related to inclusiveness and fairness with metrics like statistical parity, transparency, objectivity (also classified in data quality), and trustworthiness.

## 5 DISCUSSION

In this section, we provide our insights on the domain of data valuation and data monetisation and address the limitations of our study and the challenges currently unsolved in the literature.

As mentioned previously, we have conducted an expansive survey and tried to include as many metrics and KPIs as possible in the context of data valuation and data monetisation. Our objective is to push towards the creation of a standard framework covering every aspect of an organisation's business. However, we are also aware of the difficulty of creating such a standard and acknowledge the efforts already made by the research community to create such frameworks already (e.g. data quality frameworks [105, 132]). By focusing our study on data valuation and data monetisation, we may also have missed some metrics that could be used for such purposes.

One major challenge in the creation of a standard framework comes from the fact that many metrics seem to overlap or have very similar usages and/or definitions (e.g. timeliness vs currency). In our effort to classify all these metrics and KPIs, we have merged many of them together since they clearly represent the same concept, or opposite concepts that can be computed directly from one another. For instance, we have decided to group clarity and understandability. While some may argue they differ, we think they overall represent the same concept of preventing confusion when using data. The way we have merged similar metrics and created the taxonomy is fundamentally subjective and will certainly be subject to changes in the future. Miller et al. [105] and Schwabe et al. [132], like us, decided to merge many synonymous metrics together into their framework. Unfortunately, these overlaps in definitions and uses slow down the creation of a single standard, and have led to the creation of plethora of data quality frameworks over the years [104]. These discrepancies present a real challenge for the creation of a global data market in which every organisation can fairly evaluate the value of their data but also the value of the data they may purchase.

Another clear challenge in the domain is the creation of new metrics on a regular basis (e.g. data purity [17] in 2024). These new metrics may be necessary with advances in specialised domains. For instance, the recent advances in AI have favoured the creation of many metrics associated with the specific needs of AI: class parity, data fairness, and label noise to name a few. One major challenge for a standard framework is thus to keep up with the creation of these new metrics. This will require a constant effort to identify, normalise, and standardise the use of these new metrics. A standard framework will thus be dynamic and will incorporate quickly the new needs expressed by different industries.

With these two challenges in mind, we aim at keeping track of new metrics and will continue our effort to clarify the definitions of every one of them. Our future work will focus on how the taxonomy presented here was created and in particular, we will detail the possible links between all the metrics listed in this study and how these metrics are currently used. We will also detail how we currently use this taxonomy in a decision-making process to help stakeholders identify the most important metrics depending on their strategies and preferences.

## 6 CONCLUSION

In this study, we have conducted a vast systematic literature review for KPIs and metrics for data valuation and monetisation. We have tried to cover every aspect of an organisation's business by incorporating a large variety of metrics. We have then categorised all the metrics found into a taxonomy following the BSC approach with further subclustering, providing a better insight into the usage and utility of all these metrics.

We have tried to be as exhaustive as possible in finding all the KPIs and metrics that can cover every aspect of an organisation's business but also a variety of stakeholders (e.g., data brokers, data centres, etc.). The objective of this study was to provide a point of entry to these stakeholders into the domain of valuing and monetising their data.

In future work, we plan on detailing more the taxonomy and the metrics found in the literature to provide more context, better explanations, and example of their use. We will also show how we can use the created taxonomy with a decision-making approach for stakeholders to design data monetisation strategies by expressing their preferences in terms of BSC perspectives and metrics/KPIs to be used.

## ACKNOWLEDGMENTS

This research was partially supported by the EU's Horizon Digital, Industry, and Space program under grant agreement ID 101092989-DATAMITE. Additionally, we acknowledge Science Foundation Ireland under Grant No. 12/RC/2289 for funding the Insight Centre of Data Analytics (which is co-funded under the European Regional Development Fund).

Metrics, KPIs, and Taxonomy for Data Valuation and Monetisation - A Systematic Literature Review    000:29

Metrics, KPIs, and Taxonomy for Data Valuation and Monetisation - A Systematic Literature Review 000:31

000:32 Vyhmeister et al.
[166] Torsten Wilde, Axel Auweter, Michael K. Patterson, Hayk Shoukourian, Herbert Huber, Arndt Bode, Detlef Labrenz, and Carlo Cavazzoni. 2014. DWPE, a new data center energy-efficiency metric bridging the gap between infrastructure and workload. In *2014 International Conference on High Performance Computing & Simulation (HPCS)*. 893–901. https://doi.org/10.1109/HPCSim.2014.6903784

[167] Haocheng Xia, Jinfei Liu, Jian Lou, Zhan Qin, Kui Ren, Yang Cao, and Li Xiong. 2023. Equitable data valuation meets the right to be forgotten in model markets. *Proceedings of the VLDB Endowment* 16, 11 (2023), 3349–3362.

[168] Tao Xiaoming, Wang Yu, Peng Jieyang, Zhao Yuelin, Wang Yue, Wang Youzheng, Hu Chengsheng, and Lu Zhipeng. 2024. Data component: An innovative framework for information value metrics in the digital economy. *China Communications* 21, 5 (2024), 17–35.

[169] Anran Xu, Zhenzhe Zheng, Qinya Li, Fan Wu, and Guihai Chen. 2023. VAP: Online Data Valuation and Pricing for Machine Learning Models in Mobile Health. *IEEE Transactions on Mobile Computing* (2023).

[170] Anran Xu, Zhenzhe Zheng, Fan Wu, and Guihai Chen. 2022. Online data valuation and pricing for machine learning tasks in mobile health. In *IEEE INFOCOM 2022-IEEE Conference on Computer Communications*. IEEE, 850–859.

[171] Mehdi Yalaoui and Saida Boukhedouma. 2021. A survey on data quality: principles, taxonomies and comparison of approaches. In *2021 International Conference on Information Systems and Advanced Technologies (ICISAT)*. IEEE, 1–9.

[172] Ynvolve. 2025. Mastering Data Center Efficiency: The Power Metrics. https://ynvolve.com/all/mastering-data-center-efficiency-the-power-metrics/ Accessed: 2025-06-09.

[173] Ziba Yusufi, Simon J Preis, Daniel Kraus, Udo Kruschwitz, and Bernd Ludwig. 2022. Data Value Assessment in Semiconductor Production. (2022).

[174] Jing Zhang and Xiao-Ping Liu. 2018. Evaluation of Integrated Logistics Information System Based on Perception. In *Proceedings of the 2018 1st International Conference on Internet and e-Business*. 319–323.

[175] Y. Zhou, F. Tu, K. Sha, J. Ding, and H. Chen. 2024. A Survey on Data Quality Dimensions and Tools for Machine Learning Invited Paper. In *Proc. - IEEE Int. Conf. Artif. Intell. Test., AITest*. Institute of Electrical and Electronics Engineers Inc., 120–131. https://doi.org/10.1109/AITest62860.2024.00023 Journal Abbreviation: Proc. - IEEE Int. Conf. Artif. Intell. Test., AITest.